\documentclass{webofc}
\usepackage[varg]{txfonts}   
\usepackage{siunitx}
\usepackage{graphicx}
\usepackage{lineno}

\newcommand{\ttbar}{\ensuremath{{\text{t}\overline{\text{t}}}}\xspace} 
\newcommand{\pt}{\ensuremath{\text{p}_\text{T}^{}}\xspace}

\begin{document}
\title{Reconstructing jets in the Phase-2 upgrade of the CMS Level-1 Trigger with a seeded cone algorithm}
%
%

\author{\firstname{Sioni} \lastname{Summers}\inst{1}\fnsep\thanks{\email{sioni@cern.ch}} \and
        \firstname{Ioannis} \lastname{Bestintzanos}\inst{2,1} \and
        \firstname{Giovanni} \lastname{Petrucciani}\inst{1} \newline
        on behalf of the CMS Collaboration
}

\institute{CERN (Geneva, Switzerland) \and
           University of Ioannina (Ioannina, Greece)
}

\abstract{%
  The Phase-2 Upgrade of the CMS Level-1 Trigger (L1T) will reconstruct particles using the Particle Flow algorithm, connecting information from the tracker, muon, and calorimeter detectors, and enabling fine-grained reconstruction of high level physics objects like jets. We have developed a jet reconstruction algorithm using a cone centred on an energetic seed from these Particle Flow candidates. The implementation is designed to find up to 16 jets in each Xilinx Ultrascale+ FPGA, with a latency of less than $\qty{1}{\micro\second}$, and event throughput of $\qty{6.7}{\mega\hertz}$ to fit within the L1T system constraints. Pipelined processing enables reconstruction of jet collections with different cone sizes for little additional resource cost. The design of the algorithm also provides a platform for additional computation using the jet constituents, such as jet tagging using neural networks. We will describe the implementation, its jet reconstruction performance, computational metrics, and the developments towards jet tagging.
}
\maketitle
\vspace{-2em}
\section{Introduction}
\label{intro}

The Phase-2 Upgrade of the CMS Level-1 Trigger (L1T) \cite{tdr_l1t} will have access to new, highly-granular, data not available in the Phase-1 system.
Tracks will be reconstructed for every collision event from modules in the silicon outer tracker for charged particles with transverse momentum (\pt) above $\SI{2}{GeV}$ in the L1T Track Finder \cite{tdr_tracker}.
The new endcap high-granularity calorimeter, a sampling calorimeter with high transverse and longitudinal granularity will send clusters to the L1T \cite{tdr_hgcal}.
In addition the barrel calorimeter will provide more granular data to the L1T than during Phase-1.
The upgraded L1T has a latency restriction of $\SI{12.5}{\micro s}$ to return its select/reject decision back to the detector for event readout.
Of this total, approximately $\SI{5}{\micro s}$ are allocated to the reconstruction of charged particle tracks in the outer tracker, calorimeter clusters, and muon tracks, and around $\SI{1}{\micro s}$ to each of two `correlator' layers that are be described in Section \ref{sec:algorithm}.
Jets will be reconstructed in the second of these correlator layers, implying a $\SI{1}{\micro s}$ latency budget for the jet reconstruction.
The baseline design comprises Xilinx Ultrascale+ FPGAs and high speed optical transceivers for data transimission between devices.

The correlator subsystem of the L1T will carry out Particle Flow (PF) \cite{pflow} reconstruction and Pileup per Particle Identification (PUPPI) \cite{puppi} pileup suppression.
PF reconstructs particle candidates from the charged particle and muon tracks, and calorimeter clusters by linking compatible measurements.
PUPPI supresses the contribution from the 200 pileup (PU) vertices planned for the High Luminosity LHC (HL-LHC).
Charged PF particles are removed by PUPPI if they are not associated to the reconstructed Primary Vertex, and the \pt of neutral particles is weighted according to the local neighbourhood of other particles.
The implementation and performance of PF and PUPPI for the L1T are described in \cite{l1pfp_kreis, l1pfp_gpetrucc, l1pfp_herwig}.

Jets will be reconstructed from the PUPPI candidates in the L1T.
In the CMS High Level Trigger (HLT) and offline reconstruction, jets are reconstructed with the anti-kt \cite{anti-kt} algorithm.
This algorithm is used as the benchmark of performance in Section \ref{sec:performance}, but due to the large number of iterations required, it is deemed unsuitable for implementation within the $\SI{1}{\micro s}$ constraint.
Cone algorithms for reconstructing jets have previously been used at CMS, but are disfavoured due to lack of colinear and infra-red safety \cite{jetreview}.
However the computational simplicity makes them a good candidate for fast implementation in the online domain.

We implemented a simple jet finding algorithm using a cone centred on an energetic seed particle.
Our implementation uses the full granularity of information available from the reconstructed particles, and fits within the L1T system constraints.
The implementation provides a platform for development of further sophisticated jet processing, that will take the capabilities of the L1T beyond what is possible with the Phase-1 system.
Here we describe the concurrent reconstruction of jets with both $R=0.4$ and $R=0.8$ radius parameters: `SC4' and `SC8' respectively.
SC4 jets are used for studies with the baseline trigger selections, while the SC8 jets are under investigation for improving the performance of the L1T in particular for boosted topologies.
This may have applications to extend the physics reach of the CMS Phase-2 L1T, for example in searches for low-mass resonances decaying to boosted jets, that have required some initial- or final-state radiation in order to pass a trigger selection in previous analyses \cite{lowmass_dijet, lowmass_dijet_photon_isr}.

In addition, reconstructing jets from particles (as opposed to calorimeter tiles as in the Phase 1 system) enables further processing of the jets based on their constituents.
This has already been used for the development of b-tagging in the L1T, using a Neural Network to classify jets originating from bottom quarks using the information of each constituent \cite{sc4_btag}.

\section{Algorithm}
\label{sec:algorithm}


\subsection{Deregionizer}

Before finding jets, the data arriving into each Correlator Layer 2 FPGA are reorganised with the objective of simplifying the implementation of the jet algorithm itself. 
In the baseline design, Correlator Layer 1 comprises 36 FPGAs, divided into 6 event processors (through time multiplexing), each further divided into 6 regional processors in $(\eta, \phi)$ space.
Each Layer 1 board transmits the PUPPI candidates over 3--6 links operating at $\SI{25}{Gbs^{-1}}$ to Layer 2, depending on the detector region.
Within the `macro-region' processed by one board, the PF and PUPPI candidates are reconstructed in smaller `PF-regions' to simplify the implementation. 
The objects from each PF-region are transmitted from Correlator Layer 1 to Layer 2 across several clock cycles, and PF-regions are processed and transmitted sequentially.
The total transmission period for the PUPPI objects equates to the time multiplexing subdivision of 6 LHC bunch crossings, or $\SI{150}{ns}$, after which time the next event is transmitted.

\begin{figure*}[h]
\centering
\includegraphics[width=0.8\textwidth]{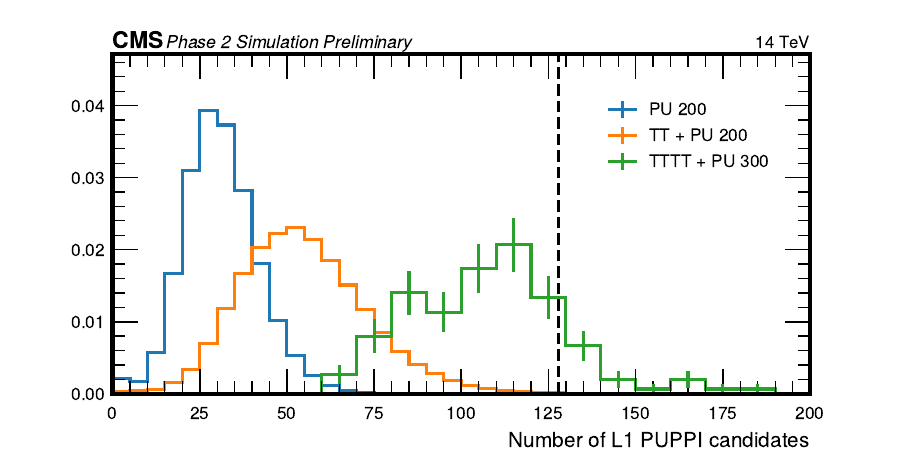}
\caption{Distributions of multiplicity of L1 PUPPI particles in different simulated signal samples. The dashed line shows the deregionizer module limit, after which particles are truncated.}
\label{fig:deregionizer_trunc}
\end{figure*}

At the receiving end in Correlator Layer 2, a `deregionizer' is placed that receives and aggregates the PUPPI candidates from each Layer 1 macro-region, and outputs one `flattened' list of particles once per event.
The deregionizer removes null objects from the input packet and aligns all valid particles to one end of the list, with no gaps.
The output of the deregionizer is a fixed-sized list, with the size chosen from balancing the number of L1T PUPPI candidates observed in Phase-2 simulations with the cost of processing those candidates in the jet finding.
Figure \ref{fig:deregionizer_trunc} show the event multiplcitiy of PUPPI candidates in different simulated processes, with the deregionizer output size of 128 particles, after which particles would be truncated.

\subsection{Jet finding}
\label{sec:jet_finding}

From the flattened list of particles, the following steps are repeated until either no particles remain in the list, or the maximum number of jets are found:
\begin{enumerate}
    \item find the particle with the highest \pt: the jet seed
    \item find particles within the jet radius, $R$, of the seed: the constituents
    \item define the jet axis as the sum over the constituents' \pt, and the \pt-weighted average of $\eta$ and $\phi$
    \item correct the jet \pt using a table of correction factors derived from matching reconstructed jets to simulated jets
    \item remove the jet constituents from the list of particles
\end{enumerate}

The algorithm is simple, considering an efficient, low latency implementation in FPGAs.
Nonetheless, using the full granularity of information from constituent particles, reconstructed from measurement in multiple subsystems, represents a development beyond the Phase-1 capability and the Phase-1-inspired algorithm described in \cite{tdr_l1t}.

\section{Implementation}
\label{sec:implementation}

As described, the simple jet algorithm was designed as such in order to reach a low latency and fit within the resource constraints of the Virtex Ultrascale+ FPGAs of the Phase 2 trigger system.
Even so, the implementation makes use of several techniques in order to fit the performance requirements.
As a design principle, we seek to maximise use of the two types of parallelism afforded by FPGAs: resource-level and pipeline-level.
Resource-level parallelism enables the concurrent execution of separate tasks, while pipeline-level parallelism enables one task to process multiple data simultaneously.

We note that the algorithm described in \ref{sec:jet_finding} has a `loop-carried dependence' \cite{hls_opt}.
That is, since the constituents of a jet are removed from the particle list before finding the next seed, the processing of iteration `$i$' cannot commence until the completion of iteration `$i-1$'.
Consequently, the total latency comprises a component proportional to the latency of the loop.
In order to minimize the latency of one loop iteration, we identify the shortest path creating the loop-carried dependence.
Referring to the algorithm steps in \ref{sec:jet_finding}, steps 1, 2, and 5 comprise this path.
As soon as the constituents of jet `$i-1$' are found and removed from the list, the seed-finding of jet `$i$' can commence simultaneously with the computation of the jet axis of jet `$i-1$' (steps 3 and 4).

The processing steps are implemented as modules in the FPGA using Xilinx High Level Synthesis (HLS) technology.
Each module is pipelined with an `initiation interval' (II) of one clock cycle, such that they can accept new data every cycle.
A control module orchestrates the injection of new event data, and the movement of data between the modules, including the iteration of the loop, implemented in VHDL.
Jets are found in order of the \pt of the seed, which does not equate to ordering by \pt of the jet, although they are correlated.
After the \pt of each jet is corrected, the jet is inserted into a module that orders them by \pt.

\begin{figure*}
    \includegraphics[width=\textwidth]{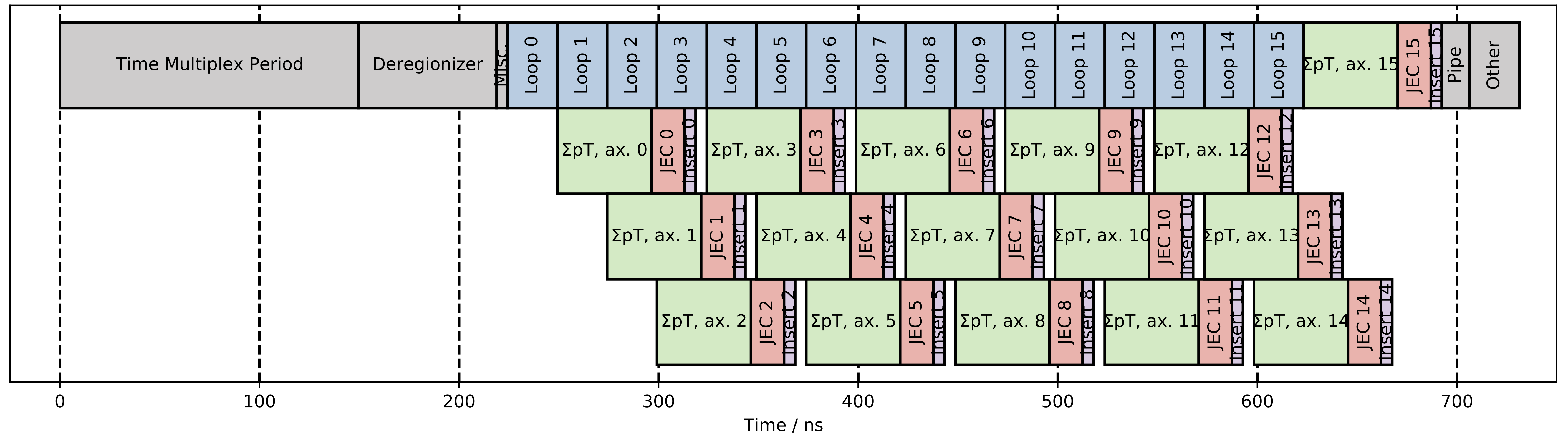}
    \caption{Processing timeline of one event, showing concurrent execution of tasks. The tasks contributing to the latency critical path are grouped along the top row.}
    \label{fig:timeline}
\end{figure*}

Figure \ref{fig:timeline} shows a timeline view of the module processing.
Each iteration of a block is executed in the same FPGA logic module (e.g. `Loop 0' and `Loop 1' refer to the same module processing different data). 
The top row of the timeline depicts the critical path of the latency.
From the start of the event at $t = 0$, the Time Multiplex Period of $\SI{150}{ns}$ and the Deregionizer latency are spent receiving and reorganising the particles from Correlator Layer 1.
At $\SI{225}{ns}$ the jet finding iteration begins, with 16 iterations of finding a seed and the in-cone constituents. 
The latency of all but one of the iterations of jet axis computation and correction is hidden due to the overlap of processing with the jet finding loop.
The total latency from the first input particle to the first output jet (FIFO) is $\SI{720}{ns}$, with an additional $\SI{138}{ns}$ to transmit the output jets serially on one link.

Given the previously described loop carried dependence, and the II=1 implementation of the processing modules, only every $8^\text{th}$ pipeline slot of the modules as shown in Figure \ref{fig:timeline} is utilised. 
This facilitates a concurrent processing of events in the same modules with pipeline parallelism.
Since events exit the Deregionizer at intervals of $\SI{150}{ns}$, and the latency of the 16 jet iterations -- the duration that an event occupies one slot in the loop pipeline -- is $\SI{400}{ns}$, up to three events will be circulating at any one time.
The control module that orchestrates the jet loop tracks the iteration of each event, sending the jets of each event into separate sorting modules.
Without this mechanism, three parallel copies of the jet algorithm would be needed to keep up with the event throughput.

With event concurrency, up to three of every eight pipeline slots are utilised.
Since more than half of the possible pipeline slots remain unused, we are able to process each event twice without significant extra resource cost.
The radius of the cone used in the constituent finding loop is configurable from the orchestration module, so each event can be processed with two different radius parameters.
We therefore produce jets reconstructed with a radius of $R=0.4$ and $R=0.8$ for each event.
As for event concurrency, the jet collections for each radius setting are sorted in separate modules.

Up to 12 reconstructed jets of each collection ($R=0.4,0.8$) are finally transmitted to the Global Trigger, which makes the final event selection decision by comparing all upstream computed high-level objects (jets, $\tau$s, electrons, muons, photons, missing transverse energy, and more) against a `menu' of conditions on which to trigger \cite{l1gt_bortolato}.

\begin{figure*}
    \centering
    \includegraphics[angle=90,width=0.8\textwidth]{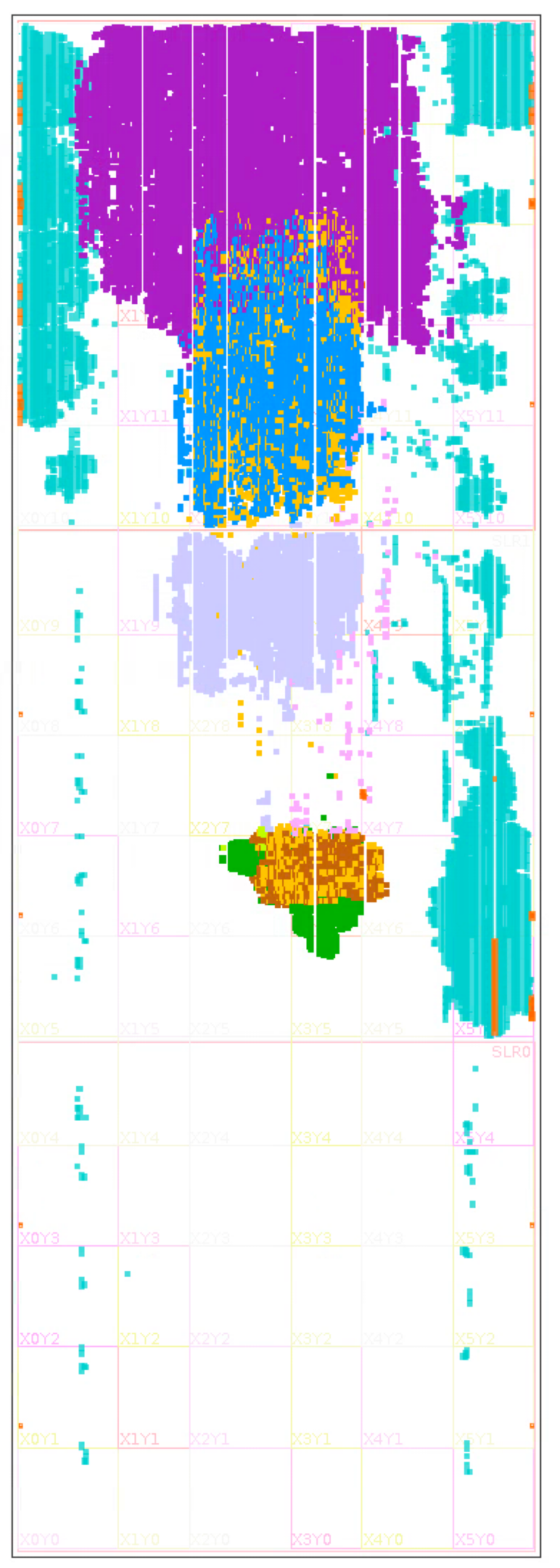}
    \caption{Floorplan of the Seeded Cone Jet algorithm in a VU9P FPGA.}
    \label{fig:floorplan}
\end{figure*}

Figure \ref{fig:floorplan} shows the floorplan of placement of jet finding modules in a VU9P FPGA on a Serenity ATCA card \cite{serenity}.
Inputs arrive on multi-gigabit transceivers in the leftmost Super Logic Region (SLR) tile.
The placement of modules is loosely constrained to flow from left to right through the floorplan as shown. 
One SLR remains completely unutilised, leaving space for the development of more sophisticated processing.

The resource usage of the algorithm and its modules in the VU9P FPGA is shown in Table \ref{tab:resources}.
The overall resource usage is very small, reaching at most $7.6 \%$ of an individual resource type, namely the DSPs.
These are used primarily for multiplications in the computation of $\Delta R$ between the seed and each particle.

\begin{table}[h]
    \centering
    \begin{tabular}{l|r|r|r|r}
         Module & LUT (\%) & FF (\%) & DSP (\%) & BRAM (\%) \\
         \hline\hline
         Total, of which: & 7.4 & 6.2 & 7.6 & 0.1 \\
         \hline
         Deregionizer & 3.5 & 2.3 & 0.0 & 0.0 \\
         Jet Finding & 3.5 & 3.6 & 7.5 & 0.0 \\
         Jet Corrections & 0.0 & 0.0 & 0.0 & 0.0 \\
         Hadronic Sums & 0.3 & 0.1 & 0.1 & 0.0 \\
    \end{tabular}
    \caption{Resource usage of the jet finding algorithm as a percentage of the Xilinx VU9P FPGA, for Look Up Tables (LUT), Flip Flops (FF), Digital Signal Multipliers (DSP) and Block RAMs (BRAM).}
    \label{tab:resources}
\end{table}
\vspace{-3em}

\section{Performance}
\label{sec:performance}

The performance of the L1T seeded-cone jet reconstruction algorithm has been evaluated on simulated LHC events of different physics processes, propagated through the simulation of the response of the CMS detector.
Where possible all L1T processing algorithms are emulated in C++ as part of the CMS experiment software framework CMSSW.
These emulators are designed to match the behaviour of the FPGA implementations at a bit-exact level.

From the emulated L1T PUPPI particles, jets are reconstructed using both the seeded-cone and anti-kt jet algorithms.
The seeded-cone implementation is fully emulated, while the anti-kt algorithm is the offline implementation in the FastJet package \cite{fastjet}.
The emulation matches the hardware implementation with bit-exact agreement on $99.9\%$ of jets.
Jet \pt corrections are derived from matching of SC4 L1T PUPPI jets to AK4 jets reconstructed from simulated particles in bins of $\eta$ and \pt, using a sample of simulated \ttbar events with 200 pileup.
In each bin a correction factor is derived that is applied to the SC4 L1T PUPPI jets.



\begin{figure*}
\begin{tabular}{cc}
\includegraphics[width=0.5\textwidth]{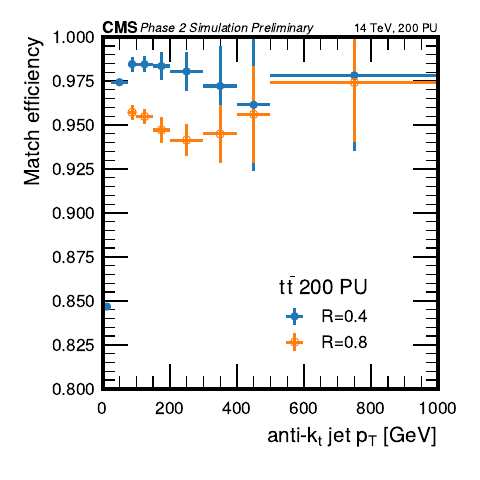} &
\includegraphics[width=0.5\textwidth]{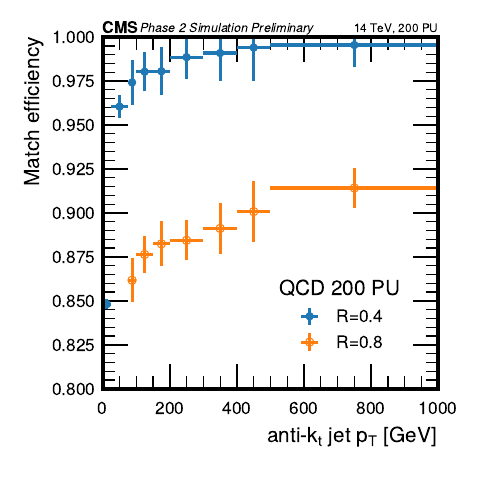} \\
\multicolumn{2}{c}{\includegraphics[width=0.5\textwidth]{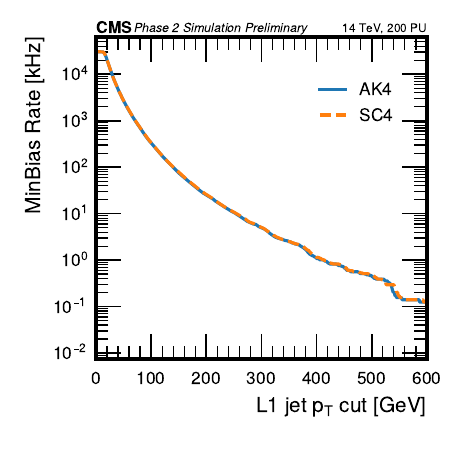}}
\end{tabular}

\caption{Top row: matching efficiency of seeded-cone jet algorithm with respect to anti-kt jets of the same radius parameters, for both $R=0.4$ and $R=0.8$, in simulated events of \ttbar (left) and QCD (right), both with 200 pileup. Bottom row: trigger rate of a selection on the highest \pt jet of each event, for seeded-cone (SC4) and anti-kt (AK4) with radius parameter $R=0.4$, in simulated `minimum bias' (MinBias) events with no hard physics process, and 200 pileup.}
\label{fig:phys_matching_rate}
\end{figure*}

The matching efficiency of seeded-cone to anti-kt jets is shown in Figure \ref{fig:phys_matching_rate} (top row), with radius parameters of $R=0.4$ and $R=0.8$ used for both algorithms.
Each anti-kt jet is considered matched when a seeded-cone jet is found within $\Delta R \leq 0.2$ and $0.8 \leq p_{T,AK}/p_{T,SC} \leq 1.2$.
We observe excellent matching of SC4 to AK4 in events with simulated QCD jets and 200 pileup.
The matching efficiency of SC4 to AK4 is very high in events of \ttbar and 200 pileup, though somewhat worse than in the QCD jets.
Matching of SC8 to AK8 is better in \ttbar events than QCD.
We understand these differences to be a consequence of the seeded-cone algorithm seeding on the single highest \pt constituent, and the presence or absence of boosted jets in the events.

Figure \ref{fig:phys_matching_rate} (bottom row) shows the rate of a trigger selection on the \pt of the leading L1T jet of each event on `minimum bias' data -- events with only pileup vertices and no hard physics process which are representative of the majority of LHC collisions.
The seeded-cone and anti-kt performance is very similar.

\begin{figure*}
\includegraphics[width=0.5\textwidth]{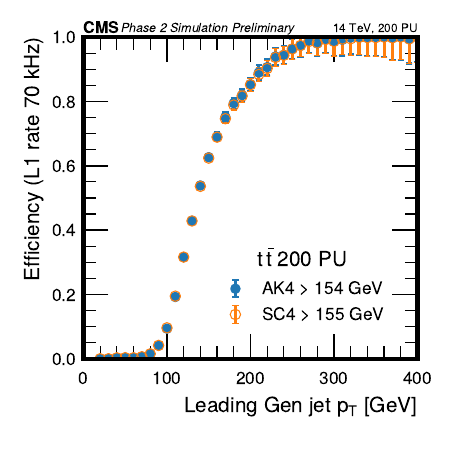}
\includegraphics[width=0.5\textwidth]{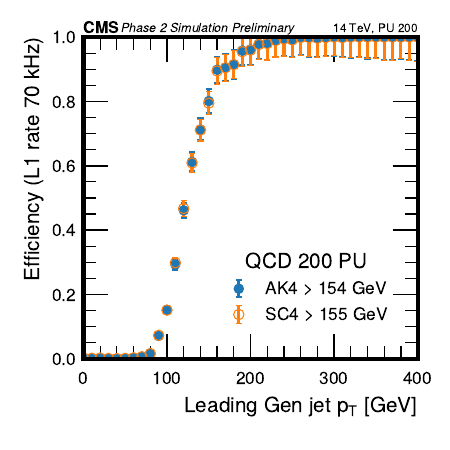}
\includegraphics[width=0.5\textwidth]{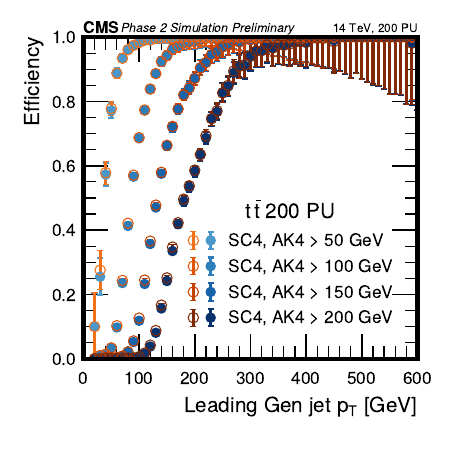}
\includegraphics[width=0.5\textwidth]{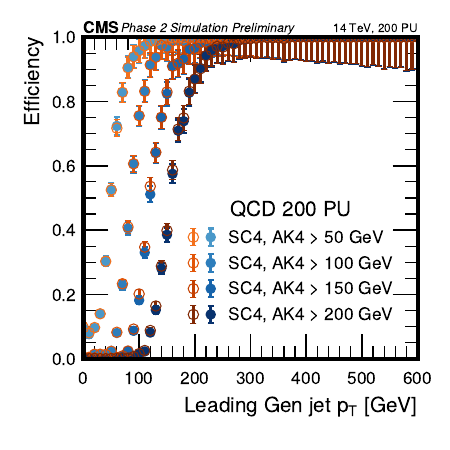}
\caption{Top: turn on curve at a fixed rate of $\SI{70}{kHz}$, for seeded-cone (SC4) and anti-kt (AK4), with radius parameter $R=0.4$, in simulated events of \ttbar (left) and QCD (right), both with 200 pileup. Bottom: turn on curves with different thresholds applied to the leading jet \pt, for SC4 and AK4 in simulated events of \ttbar (left) and QCD (right), both with 200 pileup.}
\label{fig:phys_turnons}
\end{figure*}

Figure \ref{fig:phys_turnons} shows the efficiency turn on curves for a leading jet trigger at a fixed rate of $\SI{70}{kHz}$ (top row), with the thresholds derived from Figure \ref{fig:phys_matching_rate}, and turn on curves for different thresholds from 50 to $\SI{200}{GeV}$ (bottom row).
The seeded cone performance is typically very close to the anti-kt performance in all cases.
At every threshold studied the efficiency reaches a plateau at 100\%.
The steepness of turn on varies between the different thresholds and simulated samples.
At the higher \pt thresholds the turn on in \ttbar with 200 pileup is noticeably slower than at the same threshold in QCD with 200 pileup.
The seemingly lost efficiency in \ttbar is understood to arise from cases where the system is insufficiently boosted to produce a single narrow jet that is completely captured by the R=0.4 reconstruction.
These events may still be triggered by selections on multiple jets, hadronic energy sums, or the large radius jets.

\section{Conclusions}

We have described the implementation of a seeded cone jet algorithm, a lightweight jet reconstruction designed for the tight latency constraint of the CMS Phase-2 Level-1 Trigger system.
The implementation fits within one of the Virtex Ultrascale+ FPGAs of the system, with adequate resources still available for further jet processing such as jet-tagging.

The performance of jet triggers on the Seeded Cone jets matches closely to that reached with anti-kt jets reconstructed from the same L1 PUPPI particles, showing that the simple algorithm sufficiently utilises the information available from its inputs.

In addition to reconstructing jets for use in a baseline menu complementary to the CMS Phase-1 physics program, the seeded cone jet algorithm provides a platform to extend to a rich set of jet triggers.
Tagging of b-jets using the seeded cone constituents has already been demonstrated in \cite{sc4_btag}.
Here we described the extension to jets reconstructed with a radius $R=0.8$, which we anticipate will be used to develop triggers, and possibly taggers, of boosted jet topologies.

\vspace{{-1em}}
\bibliography{refs}

\end{document}